\documentclass[aps,amsfonts,prx,reprint,twocolumn,superscriptaddress,showpacs]{revtex4-1}

\usepackage[dvips]{graphicx}
\usepackage{amssymb,amsfonts,amsmath,bm}
\usepackage[usenames]{color}
\usepackage{xcolor}

\usepackage{refstyle}
\usepackage{mathrsfs}
\usepackage{esint}
\usepackage{cancel}
\usepackage[normalem]{ulem}
\usepackage[caption=false]{subfig}
\usepackage[unicode=true,pdfusetitle,bookmarks=true,
bookmarksnumbered=false,bookmarksopen=false,breaklinks=false,
pdfborder={0 0 0},backref=false,colorlinks=true,citecolor=red]
{hyperref}
\usepackage{cleveref}
\providecommand\eqref[1]{\ref{eq:#1}}
\renewcommand\b[1]{{\bf  #1}}
\renewcommand\vec[1]{\boldsymbol{#1}}
\renewcommand\phi{\varphi}

\newcommand\del{\nabla}
\newcommand\dd{\mathrm{d}}

\allowdisplaybreaks[3]

\newcommand{\PT}{{\cal{{PT}}}}

\usepackage{wrapfig, framed}

\newcommand{\mcm}[1]{\textcolor{black}{#1}}
\renewcommand{\ss}[1]{\textcolor{black}{#1}}

\newcommand{\mjb}[1]{{\textcolor{black}{#1}}}

\begin{document}

\title{Topological active matter}
\author{Suraj Shankar}
\affiliation{Department of Physics, Harvard University, Cambridge, MA 02318, USA}
\author{Anton Souslov}
\affiliation{Department of Physics, University of Bath, Claverton Down, Bath BA2 7AY, UK}
\author{Mark J.~Bowick}
\affiliation{Kavli  Institute  of  Theoretical  Physics,  University  of  California  Santa  Barbara,  Santa  Barbara,  CA  93106}
\author{M.~Cristina Marchetti}
\affiliation{Department of Physics, University of California Santa Barbara, Santa Barbara, CA 93106, USA}
\author{Vincenzo Vitelli}
\affiliation{James Franck Institute, The University of Chicago, Chicago, IL 60637, USA}
\affiliation{Department of Physics, The University of Chicago, Chicago, IL 60637, USA}
\affiliation{Kadanoff Center for Theoretical Physics, The University of Chicago, Chicago, IL 60637, USA}
\date{\today}

\begin{abstract}
\ss{Active matter encompasses different nonequilibrium systems in which individual constituents convert energy into non-conservative forces or motion at the microscale. This review provides an elementary introduction to the role of topology in active matter through experimentally relevant examples. Here, the focus lies on topological defects and topologically protected edge modes with an emphasis on the distinctive properties they acquire in active media. These paradigmatic examples represent two physically distinct classes of phenomena whose robustness can be traced to a common mathematical origin: the presence of topological invariants. These invariants are typically integer numbers that cannot be changed by continuous deformations of the relevant order parameters or physical parameters of the underlying medium. We first explain} the mechanisms whereby topological defects self propel and proliferate in active nematics, leading to collective states which can be manipulated by geometry and patterning. \ss{Possible implications for active microfluidics and biological tissues are presented.} We then illustrate how the propagation of waves in active fluids and solids is affected by the presence of topological invariants characterizing their dispersion relations. \ss{We discuss the relevance of these ideas for the design of robotic metamaterials and the properties of active granular and colloidal systems. Open theoretical and experimental challenges are presented as future research prospects.}
\end{abstract}
\maketitle

\ss{
Active media are out-of-equilibrium systems composed of individual components that convert energy into non-conservative forces and motion at the microscale~\cite{ramaswamy2010mechanics,marchetti2013hydrodynamics,ramaswamy2017active}. Examples include self-propelled particles (e.g., micron-sized colloids powered by chemical reactions)~\cite{palacci2013living,bricard2013emergence}, biofilaments with molecular motors~\cite{schaller2010polar,sanchez2012spontaneous} and many biological systems ranging from epithelial tissues and bacterial colonies to bird flocks~\cite{needleman2017active,cavagna2014bird,trepat2018mesoscale}. Describing these media using continuum equations, as you would for passive fluids or solids, requires a careful re-examination of the symmetries and conservation laws that are present (or absent!) in each active system~\cite{marchetti2013hydrodynamics}. 
The most obvious example is the conservation of energy, which is manifestly violated by the presence of molecular motors or other mechanisms of energy transduction at the microscale that power self-sustained flows and active stresses. The energy injected at the microscale is responsible for morphological features and transport properties absent at equilibrium that are robust against certain perturbations. Some of these robust features can be analyzed using a branch of mathematics called topology.}

\ss{
Topology describes properties of objects that are preserved under continuous deformations of their shapes.
A common example is the smooth deformation of a doughnut into a mug (with a handle). During this transformation, the shape of the object changes but the number of handles, $\nu$---seen as holes from a three-dimensional viewpoint---is preserved, as long as no tear is generated. The integer $\nu$, called the genus, is an example of a topological invariant. In condensed matter physics, a prime application of topological invariants is in the characterization of
topological defects, which are particle-like objects that describe global deformations of an ordered medium~\cite{mermin1979topological,alexander2012colloquium,nelson2002defects}. As an example, consider a vortex (Box 1) in a two-dimensional (2D) vector field that represents, for instance,
%(e.g. it is an order parameter for) 
the local orientation of elongated molecules lying in the plane. While the location of an isolated vortex can be parameterized by the position of its center, its presence disrupts molecular alignment throughout space.
}

\ss{
The global character of a vortex is captured mathematically by the definition of its winding number. 
Start by introducing the field $\theta(\b{r})$ that describes the local angle that the molecules (i.e., the vector order parameter) make with respect to a fixed direction in the plane. The winding number, $\nu$, is then an integer that tracks the cumulative change in $\theta(\b{r})$ along \emph{any} path enclosing the vortex:
\begin{equation}
\nu = \dfrac{1}{2\pi}\oint\vec{\del}\theta\cdot\dd\b{l}\;.
\label{nu}
\end{equation}
Very much like the number of handles on a surface is preserved under continuous deformations of its shape, the integer $\nu$ is also a topological invariant. The only difference is that in this case, $\nu$ is preserved under continuous deformations of the vector order parameter. As a result, a vortex characterized by the non-vanishing winding number $\nu=1$ is said to be topologically stable or robust: it cannot be made to disappear ($\nu \rightarrow 0$) unless annihilated with a topological defect characterized by a winding number of opposite sign, i.e., an anti-vortex (Box 1). Note that if the line integral in Eq.~(\ref{nu}) is evaluated along a path encircling multiple vortices and anti-vortices, $\nu$ measures the \emph{net} number of topological defects (to which a positive or negative sign is assigned depending on their individual winding numbers).
}

\ss{
A similar mathematical mechanism ensures the robustness of so-called chiral edge modes in topological insulators~\cite{kohmoto1985topological,hasan2010colloquium,qi2011topological,FRUCHART2013}. A defining feature of these wave modes is that they propagate unidirectionally along sample boundaries without experiencing any backscattering even if they encounter sharp corners or obstacles on their way~\cite{halperin1982quantized,moore2010birth}.
This property can be traced to the presence of a topological invariant called a Chern number~\cite{nakahara2003geometry}. This integer, that we once again denote by $\nu$ to emphasize the analogy with the previous case, measures the \emph{net} number of unidirectional, i.e., chiral, edge modes present (to which a positive or negative sign is assigned depending on whether the modes propagate clockwise or counterclockwise). These modes are said to be topologically robust, much like the vortices discussed above. In this case, the Chern number (and hence the very presence of an edge mode) is preserved under continuous changes in the {\it physical} parameters of the material (rather than under continuous deformations of the {\it order} parameter). On a more technical level, we shall see that topological invariants, such as Chern numbers, can also be viewed as winding numbers encircling topological defects, but in \emph{wave-vector} space.
}

\ss{
Topological defects and topological protected modes are intertwined~\cite{ran2009one,teo2010topological,jurivcic2012universal,paulose2015topological}. In addition to boundaries, topological modes can be localized around defects such as vortices, dislocations or domain walls. Topological defects~\cite{mermin1979topological,chaikin2000principles,nelson2002defects} and topologically protected modes~\cite{zhang2018topological,huber2016topological,khanikaev2017two,mao2018maxwell,ozawa2019topological,ma2019topological} occur in a variety of physical contexts, but in active media they acquire distinctive properties, which are the focus of this review.
}

\ss{
Since active media display non-conservative stresses and spontaneous flows, they naturally break key symmetries that passive materials would possess~\cite{marchetti2013hydrodynamics,ramaswamy2017active}. For example, active media can violate reciprocity~\cite{nassar2020nonreciprocity,fruchart2020phase}, defined as the symmetry between perturbation and response, as well as detailed balance~\cite{seifert2019stochastic,gnesotto2018broken}, defined as the symmetry between the past and the future within the dynamics of microscopic processes.
%as well as detailed balance, that is the microscopic expression of time-reversal invariance: stipulates that each process is balanced by its reverse. 
Non-equilibrium violations of either of these principles introduces an arrow of time and enables the presence of non-vanishing currents (of energy, momentum, or particles) in the steady state of active systems. In turn, these currents endow topological modes and defects with properties absent at equilibrium
and, in some cases, enable their very existence. In order to illustrate this point, consider the following two examples.
}

\ss{
First, topological defects (which normally cost a large elastic energy~\cite{chaikin2000principles}) can unbind and proliferate even at zero temperature if active stresses are present~\cite{shankar2018defect,shankar2019hydrodynamics}. This unbinding occurs in active nematics~\cite{ramaswamy2003active,doostmohammadi2018active}, which are liquid-crystalline media composed, for instance, of cytoskeletal filaments driven by molecular motors~\cite{sanchez2012spontaneous,kumar2018tunable}. Moreover, the resulting topological defects can move by themselves, \ss{i.e., even in the absence of external forces or fields~\cite{narayan2007long,sanchez2012spontaneous,giomi2013defect}. This spontaneous ``self-propulsion'' depends} on the winding number of the defects~\cite{giomi2013defect} and ultimately originates from the presence of non-conservative internal forces generated by ATP-powered molecular motors on the microscale.
}

\ss{
Second, chiral edge modes occur in mechanical, optical, and electronic systems~\cite{hasan2010colloquium,zhang2018topological,huber2016topological,khanikaev2017two,mao2018maxwell,ozawa2019topological,ma2019topological}, but in the simplest situation, external (e.g., magnetic) fields are typically required to break the symmetry between left and right-moving waves. A simple way to break this symmetry in mechanics is to build a lattice of circulators~\cite{khanikaev2015topologically,yang2015topological}. A circulator is a ring in which air is constantly moved by a fan, for example clockwise. Experiments~\cite{khanikaev2015topologically,yang2015topological} reveal that when many such circulators are assembled into a lattice, the sound waves propagating on top of the flowing fluid are topologically protected. If one now fills the rings with self-propelled particles (instead of a passive fluid like air), the required flow arises spontaneously without motorizing the circulators: the microscopic particles themselves are motorized~\cite{souslov2017topological}! 
}

\ss{
In the remainder of this review, we build on the examples considered above and provide an introduction to the theory of topological active matter as well as a survey of its rich experimental ramifications.
}

\begin{figure*}
    \centering
    \includegraphics[width=\textwidth]{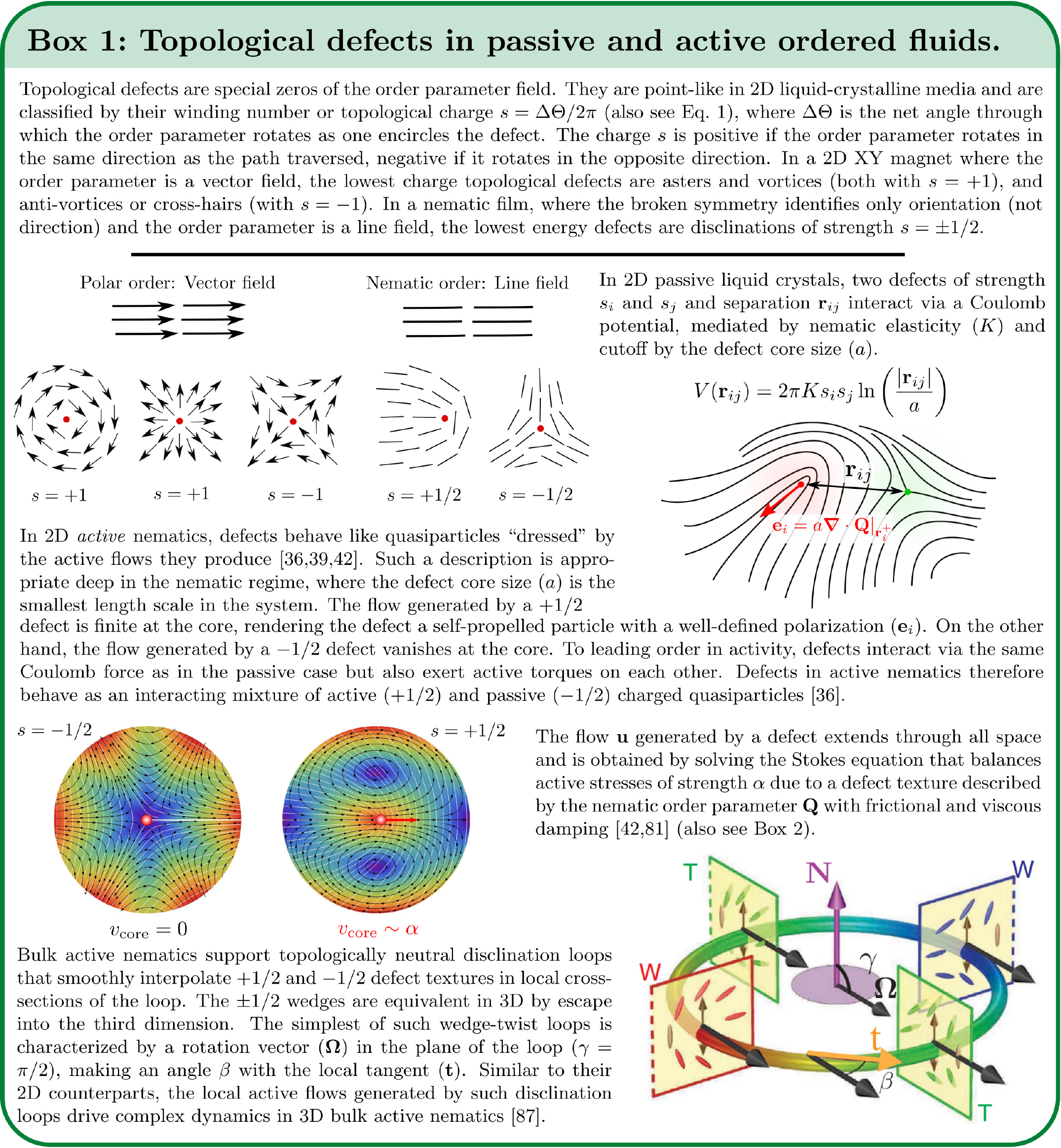}
    \captionsetup{labelformat=empty}
\end{figure*}

\section{Topological defects in active matter}

\ss{In passive materials, topological defects are inevitably formed in quenches from the disordered into the ordered state or when order is frustrated by curvature, external fields, or boundary conditions. They constitute elementary excitations of the homogeneous ordered state and their statistical mechanics offers a picture dual to that of the familiar order parameter~\cite{nelson2002defects}. Order-disorder transitions in many passive two-dimensional systems, including superfluid and superconducting films, crystalline layers and 2D nematics, are controlled by the unbinding of topological defects through the celebrated Berezinskii-Kosterlitz-Thouless mechanism (BKT)~\cite{berezinskii1971destruction,berezinskii1972destruction,kosterlitz1973ordering,kosterlitz1974critical,kosterlitz2016kosterlitz}. The BKT theory reveals a distinctive universality class of defect-induced continuous phase transitions at equilibrium~\cite{kosterlitz2016kosterlitz}. The theory relies on the mapping of the statistical physics of point defects onto a gas of interacting Coulomb charges. At low temperature, opposite-sign defect pairs are bound by Coulomb-like attraction and the state remains ordered. Above a critical temperature, entropic effects overcome energetic attraction resulting in defect unbinding, which destroys the ordered state~\cite{nelson2002defects,chaikin2000principles}.}

Just as in equilibrium, defects play an important role in the spatiotemporal dynamics and relaxation of \emph{active} ordered phases.
\ss{We focus here on orientationally ordered fluids which can host point-like defects in two dimensions (2D), and both point- (monopoles, also called asters) and line-like defects (disclinations) in three dimensions (3D)~\cite{mermin1979topological,alexander2012colloquium}.}
%In this section, we focus primarily on 2D active fluids with orientational order. 
Collections of polar active particles  with aligning interactions, akin to ``flying spins,'' can order in states of collective motion where the order parameter is the mean velocity of the flock (see Box 2). In these \emph{polar} active fluids, also called Toner-Tu fluids~\cite{Vicsek1995,toner1995long,toner1998flocks,toner2005hydrodynamics}, the ordered state is a state of mean motion that spontaneously breaks rotational symmetry and time-reversal symmetry (TRS) on a global scale.  In contrast, active nematics~\cite{ramaswamy2003active,doostmohammadi2018active} are composed of head-tail symmetric rod-like entities that exert internally generated stresses on their surroundings and organize in states of \emph{apolar} orientational order. These fluids display no net motion on average, but combine the rich rheology of liquid crystals with active driving. \ss{Box 2 describes some commonly used continuum models to describe such systems.}

Active ordered fluids have been assembled from a variety of soft materials~\cite{schaller2010polar,ndlec1997self,surrey2001physical,bricard2013emergence,sanchez2012spontaneous}. \mcm{The orientationally ordered state becomes unstable on length scales $L$ when active stresses of order $|\alpha|$ (see Box 2) exceed elastic stresses $\sim K/L^2$~\cite{simha2002hydrodynamic}, with $K$ the liquid-crystalline stiffness.  Bulk ordered states  ($L\rightarrow\infty$) are therefore unstable for any amount of activity, but can be stabilized by confinement~\cite{voituriez2005spontaneous} as well as by friction with a substrate~\cite{duclos2018spontaneous} below a critical value of activity. At large activity, all liquid-crystalline active fluids display turbulent-like dynamics with a proliferation of}  topological defects. Polar fluids with vectorial order parameters exhibit defects such as vortices and asters characterized by an integer topological charge. \ss{In contrast, the inversion symmetry of nematic order means that the order parameter maps onto itself after a winding of only $180^\circ$ around the defect core---\mjb{the small, roughly circular, region of space of radius $a$ surrounding the defect center where the nematic order parameter vanishes.} This allows defects with half-integer charge known as disclinations (see Box 2). Since the energetic cost of a defect is proportional to the square of its charge, $\pm1/2$ disclinations are the lowest energy defects in nematics; they are not allowed in polar (vector) fluids. The presence of $\pm1/2$ defects therefore provides a fingerprint of the apolar nature of the broken symmetry and a criterion for distinguishing nematic from polar order.}  

Asters and vortices have been observed in high-density motility assays~\cite{sumino2012large,schaller2013topological}, confined and membrane-bound reconstituted cortical extracts~\cite{ndlec1997self,surrey2001physical,koster2016actomyosin}, and suspensions of colloidal rollers motorized by an electric field~\cite{bricard2015emergent}. Nematic disclinations have  been identified in diverse systems, including cytoskeletal filament-based active nematic suspensions~\cite{sanchez2012spontaneous,keber2014topology,ellis2018curvature,kumar2018tunable}, collections of living cells~\cite{gruler1999nematic,kemkemer2000elastic,zhou2014living,duclos2017topological,blanch2018turbulent,saw2017topological,kawaguchi2017topological}, \mcm{and even multicellular organisms~\cite{maroudas2020topological}}.
In recent years topological defects have been the focus of intense research, particularly in active nematics~\cite{doostmohammadi2018active}, which exhibit self-sustained and spatiotemporally chaotic large-scale flows accompanied by the spontaneous proliferation of topological defects. It is then natural to ask if the chaotic dynamical state of active nematics can be understood from the perspective of defect unbinding, and if the equilibrium BKT transition has an analogue in the active realm. 

In both passive and active liquid crystals, distortions of orientational order generate flows, and flows in turn deform the ordered state. The key distinction between passive and active systems is that in the former, flows and deformations are generated by applying external fields or via boundary conditions. Such flows are transient: the system always relaxes to the equilibrium (ordered) state upon removal of perturbations or constraints. In active systems, in contrast, flows and deformations are internally generated and self-sustained. In 2D, bulk active nematics are generically unstable to bend or splay deformations generated by local active fluctuations~\cite{simha2002hydrodynamic}. As these distortions grow in time, they generate flows that further enhance the deformations, ultimately generating unbound pairs of topological defects. Such spontaneously generated defects are themselves strong distortions of orientational order and yield distinctive flow patterns~\cite{giomi2013defect,giomi2015geometry}. This intimate connection
%, almost a `duality,' 
between defects and flows, offers the opportunity to direct and localize nonlinear active flows by controlling the dynamics of topological defects~\cite{shankar2019hydrodynamics,ross2019controlling,zhang2019structuring}. Recent work has even suggested that biological systems may exploit this connection between structure and dynamics and use defects to localize stresses and perform specific biological functions~\cite{kawaguchi2017topological,saw2017topological,copenhagen2020topological,maroudas2020topological,meacock2021bacteria,yaman2019emergence,bacsaran2020formation}. These subjects will be the focus of the rest of this section.

\subsection{Active defects spontaneously move}
One of the distinguishing aspects of defects in active matter is their capacity for spontaneous and autonomous motion. The large distortions of the order parameter around a topological defect generate local active flows, whose symmetry and profile is controlled by
%stresses, which can lead to a local directed current due to the absence of detailed balance. The nature and symmetry of the local currents depends on 
the defect geometry. In particular, in 2D active nematics, the comet-like $+1/2$ defect (see Box 1) generates a flow that is finite at the defect core. The $+1/2$ defect then rides along with the flow it generates, behaving like a self-propelled polar particle~\cite{giomi2013defect,pismen2013dynamics}. In contrast, the flow generated by a $-1/2$ defect  vanishes at the defect core due to the defect's threefold symmetry (see Box 1), and to leading order in activity this defect remains a ``passive'' particle.  The direction of motion of the $+1/2$ defect is determined by its local orientation or polarity and by the sign of the active forcing. \ss{Active stresses that cause material extension along the ordering axis are called ``extensile,'' while those that contract along the same axis are called ``contractile.''} In extensile fluids, $+1/2$ defects actively propel themselves along the head of the comet, while in contractile systems the active propulsion is directed along the comet's tail.  If chiral active stresses are present, $+1/2$ disclinations self-propel instead at an angle relative to their polarity~\cite{maitra2019spontaneous,hoffmann2020chiral}, and intrinsic active spinning renders defect trajectories circular~\cite{maitra2019spontaneous}. In polar fluids, charge $+1$ spiral vortices undergo spontaneous rotation~\cite{kruse2004asters} due to the chirality of their spiral texture. Other low charge defects, such as circularly symmetric asters and antivortices, perform neither rotational nor translational spontaneous motion. In contrast to defect motion in externally driven systems, the motion of active defects is dictated by the local \emph{geometry} of the defect itself and not by a fixed external field, as would be the case, for example, for driven vortices in a superconducting film. This profound distinction leads to a plethora of phenomena characteristic of active matter.

Defects can also occur in 3D active suspensions, but they are less well understood. As in 2D, bulk active nematics in 3D are susceptible to a generic hydrodynamic instability~\cite{simha2002hydrodynamic} that spontaneously generates individual charge-neutral disclination loops~\cite{shendruk2018twist,vcopar2019topology,duclos2020topological,binysh2020three} (see Box 1). Unlike in 2D, where topological charge conservation constrains point defects to be created and annihilated in pairs of opposite charge, in 3D \emph{charge-neutral} disclination loops~\cite{alexander2012colloquium} can be nucleated on their own. The active flows caused by the director distortion around such a disclination loop cause it to stretch, twist, and buckle~\cite{binysh2020three}. The complex configurational dynamics of these loops combined with topological reconnections leads to chaotic 3D flows which, recently, have been observed in simulations~\cite{shendruk2018twist,vcopar2019topology} and experiments~\cite{duclos2020topological}. Recent theoretical work has also extended similar ideas to chiral phases such as bulk active cholesterics~\cite{whitfield2017hydrodynamic,metselaar2019topological,carenza2019rotation}. \ss{Here, $\lambda$-lines, which are defects in the cholesteric pitch with no director singularity at their core~\cite{whitfield2017hydrodynamic} and other nonsingular topological textures called half-skyrmions or merons~\cite{metselaar2019topological} can sustain coherent rotations and steady defect patterns along with spatiotemporal chaotic flows, even though these defects are not self-propelled.}
\nocite{cates2015motility}
\nocite{marchetti2016minimal}
\begin{figure*}
    \centering
    \includegraphics[width=\textwidth]{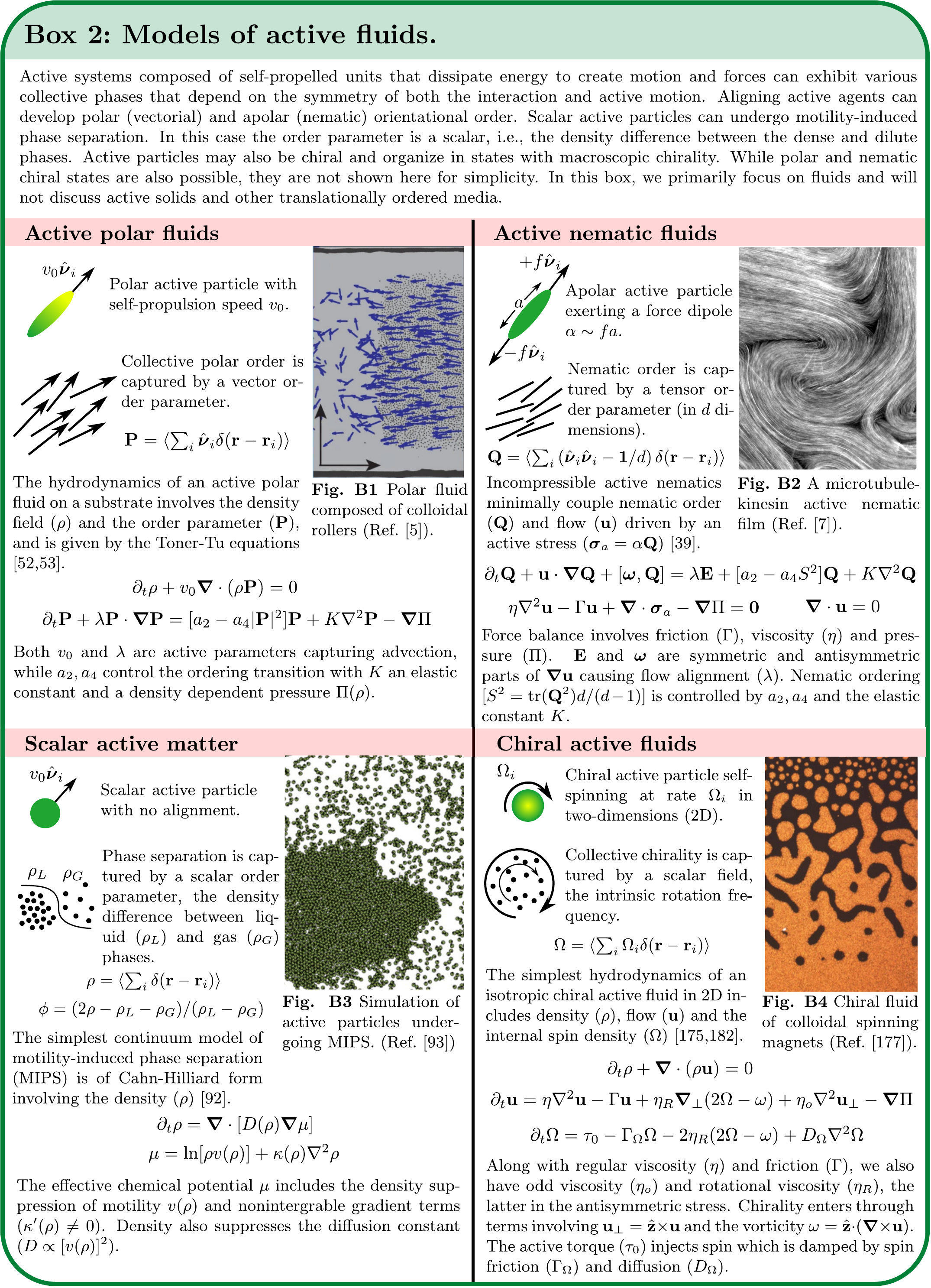}
    \captionsetup{labelformat=empty}
\end{figure*}

\subsection{Defect dynamics, unbinding, and ordering out of equilibrium}
From a fundamental point of view, defect motility raises intriguing possibilities for phase transitions. The consequences of  defect self-propulsion have been explored predominantly in 2D active nematics~\cite{thampi2013velocity,pismen2013dynamics,giomi2013defect,giomi2015geometry,shankar2018defect,shankar2019hydrodynamics}. The defect-driven chaotic flows lead to a state dubbed ``active turbulence''~\cite{wensink2012meso,giomi2015geometry}, which is characterized by vorticity and shear flows on a typical length scale that controls the mean defect separation and is set by a balance of active and elastic stresses~\cite{zhou2014living,giomi2015geometry,hemingway2016correlation,guillamat2017taming,lemma2019statistical}. \mcm{The phenomenology and scaling properties of active turbulence  have been detailed in recent reviews~\cite{doostmohammadi2018active,alert2021active}}. Here we focus  on its topological aspects and sketch the physical arguments underpinning the unbinding of active defects~\cite{shankar2018defect}.

At equilibrium, defects in 2D behave as point charges that interact via a Coulomb potential. This pair interaction is mediated by the underlying elasticity of the ordered fluid and scales as $K\ln(r/a)$, where $K$ is an elastic stiffness, $r$ is the pair separation and $a$ the defect core size (see Box 2). 
%The equilibrium BKT transition then occurs at a finite temperature beyond which entropic effects overwhelm the elastic energy, allowing defect pairs to unbind and proliferate, thereby disordering the system.
In a 2D \emph{active} nematic, an isolated $\pm1/2$ disclination pair continues to experience the passive attractive force $\sim K/r$ from elasticity, but in addition, for certain configurations, \mcm{such as the one shown in Box 1 }, the $+1/2$ defect can propel itself with speed $\sim v_0$ away from the $-1/2$ defect. The balance of elastic and active forces sets a length scale $r_c\sim K/v_0$, beyond which activity always rips apart the defect pair, causing it to inevitably unbind. This simple picture is spoiled by the fact that $+1/2$ defects do not travel in a straight line, rather their  direction of motion is affected by rotational noise and changes in the local nematic structure. Unlike the motion of charges driven by an externally applied electric field, the motility of $+1/2$ disclinations is determined by their local geometry. As a result, the $+1/2$ defect has a finite \emph{persistence length} $\ell_p=v_0\tau_R$  beyond which its motion is not ballistic, where $\tau_R^{-1}$ controls the rate of rotational noise from active processes. Upon comparing this persistence length to the length scale $r_c$ where pair attraction and defect propulsion balance, we get a simple criterion for active defect unbinding~\cite{shankar2018defect}. Defect pairs unbind if $\ell_p>r_c$. Conversely, when $\ell_p<r_c$, the $+1/2$ defect changes its direction of motion before overcoming Coulomb attraction, resulting in a local change of the nematic texture that allows the pair to remain bound. It is interesting to note that rotational noise here \emph{stabilizes} the quasi-ordered nematic phase below a finite activity threshold by disrupting the persistent motion of the $+1/2$ defect. This order-from-disorder mechanism highlights the difference between active and driven defects: the latter would inevitably unbind under the action of any external field as nothing disrupts their straight line motion.

Above a critical value of activity, the BKT-like unbinding yields an interacting gas of unbound and swarming defects that provides a useful picture to describe the state of active turbulence. Several models of varying complexity have been proposed for the dynamics of active defects~\cite{pismen2013dynamics,giomi2013defect,keber2014topology,khoromskaia2017vortex,cortese2018pair,shankar2018defect,tang2019theory,zhang2020dynamics,vafa2020multi}. While details differ, the basic physics is the same: the unbound defect gas is a mixture of self-propelled (the $+1/2$ defects) and passive (the $-1/2$ defect) charged particles interacting via  Coulomb forces. Recent work has also emphasized the non-reciprocal nature of active defect interactions~\cite{maitra2020chiral,vafa2020multi}\ss{, although its detailed consequences remain to be explored}. Coarse-graining over many defects allows a hydrodynamic description of the active defect gas~\cite{shankar2019hydrodynamics,angheluta2021role}, along the lines of previous classic works in the context of superfluid vortices~\cite{ambegaokar1980dynamics} and 2D crystal melting~\cite{zippelius1980dynamics}. Importantly, a hydrodynamic treatment of defects offers a theoretical handle on the strongly interacting many-body dynamics of this far-from-equilibrium system. As a crucial ingredient, this description includes a polarization field that captures the average orientation of a collection of active $+1/2$ defects. This field accounts not only for self-propulsion actively driving material flow, but being a vector, the orientation also experiences active torques. When activity is sufficiently strong, \mcm{and viscous stresses are negligible compared to frictional dissipation with a substrate}, one finds that the $+1/2$ defects spontaneously condense into a polar-ordered collectively moving state---a defect flock~\cite{shankar2019hydrodynamics}. The fleeting defects constantly turn over due to creation and annihilation events, but polar order persists for infinitely longer than the individual defect lifetime. 
Heuristically, such a state arises when the underlying nematic elasticity is too slow to relax the distortion created in the wake of an unbinding defect pair. Similar defect-ordered states have been observed previously in simulations, either with polar ordering~\cite{decamp2015orientational,putzig2016instabilities,srivastava2016negative,patelli2019understanding} or defect lattices~\cite{doostmohammadi2016stabilization}. Experiments on microtubule-kinesin based active nematic films have reported an extraordinary \emph{nematic} ordered defect state~\cite{decamp2015orientational}. While continuum simulations recover largely transient nematic defect ordering~\cite{oza2016antipolar,srivastava2016negative}, this observation continues to be a theoretical puzzle. Recent work suggests elastic torques may play a role in antipolar ordering of defects~\cite{pearce2020scale,thijssen2020large}.

Finally, vortex unbinding has been shown to also play a role in disordered polar active fluids~\cite{chardac2020meandering}. Here, active flows conspire with quenched obstacles to realize a dynamical vortex glass that can be rationalized through an effective Kosterlitz-Thouless-like argument in analogy with dirty superconductors. Generalizing similar ideas to active matter in heterogeneous environments~\cite{bechinger2016active} remains an open question.

\subsection{Active defects under confinement}
\begin{figure*}[]
    \center{\includegraphics[width=\textwidth]{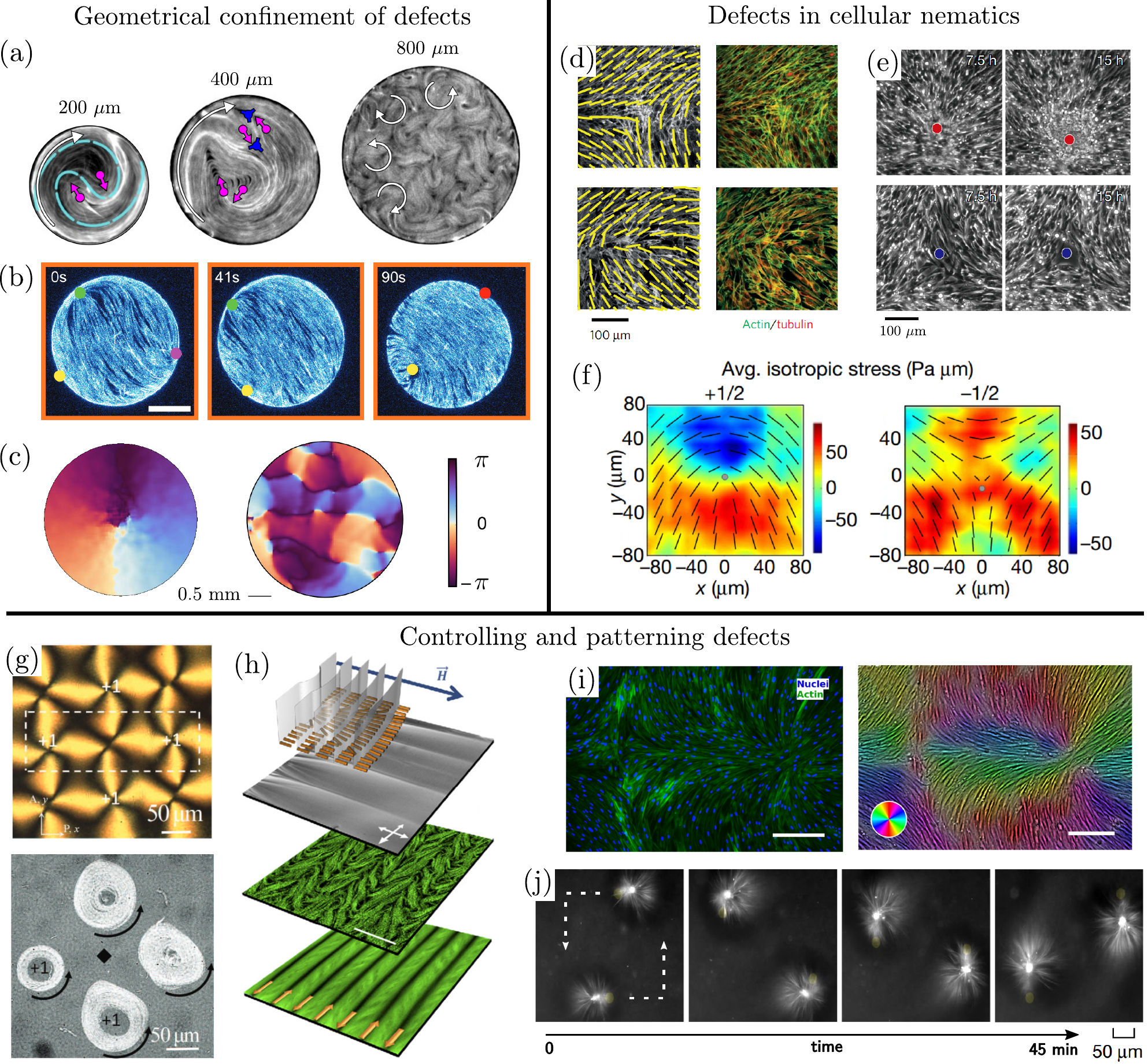}}
    \caption{\emph{Geometrical confinement of defects}: (a) Circulating $+1/2$ defects in a microtubule-kinesin active nematic film confined to a disc ($200~\mu$m), with more pairs of defects unbinding to create turbulent flows for large disc diameters ($800~\mu$m)~\cite{opathalage2019self}; (b) The same system now condensed onto a spherical vesicle (scale bar, $20~\mu$m) exhibits periodic oscillations of the four $+1/2$ defects present~\cite{keber2014topology}; (c) Schlieren texture of a colloidal polar fluid in a disc geometry displaying a system-spanning vortex (left) and a pattern of pinned vortex-antivortex pairs (right) induced by randomly located obstacles~\cite{chardac2020meandering}. \emph{Defects in cellular nematics}: (d) Active $\pm1/2$ disclinations have been identified in aligned populations of spindle-shaped mouse fibroblasts~\cite{duclos2017topological}; (e) Cells accumulate (deplete) at the cores of $+1/2$ ($-1/2$) defects, the former seeding mound formation in dense monolayers of neural progenitor cells~\cite{kawaguchi2017topological}; (f) Similar defects in epithelial monolayers of MDCK cells generate large compressive stresses (in blue) only at the head of $+1/2$ disclinations that locally trigger cell extrusion and apoptosis~\cite{saw2017topological}. \emph{Controlling and patterning defects}: (g) Directing \ss{vorticial} polar flows of bacteria dispersed in a nontoxic liquid crystal patterned with a \ss{periodic array of alternating $\pm1$} defects~\cite{peng2016command}; \ss{(h) Passive bulk smectic aligned by an external magnetic field couples hydrodynamically to an adjacent microtubule-kinesin active nematic film, forcing defects to orient and flow along alternating shear bands (scale bar, $100~\mu$m)~\cite{guillamat2016control}; (i) Patterning a tissue of human fibroblast cells (HDF) growing on a liquid crystal elastomer with a predesigned texture of $\pm1$ defects (scale bar, $300~\mu$m)~\cite{turiv2020topology}; (j) Time-series of dynamically assembled asters transported along a predetermined trajectory within a bulk microtubule-kinesin suspension, by local light activation~\cite{ross2019controlling}.}}
    \label{fig:Fig1}
\end{figure*}
The direct connection between defects and active flows  offers new avenues for rectifying and controlling the spontaneous chaotic dynamics of active nematics, paving the way to the application of these system to the development of active microfluidic devices~\cite{Green}.  Active nematic films built by depleting microtubule bundles and kinesin motor complexes onto an oil-water interface~\cite{sanchez2012spontaneous} have emerged as a versatile platform to manipulate defects by controlling the structure and materials properties of the supporting interface. One simple approach is by tuning the viscosity of the oil layer that hydrodynamically constrains the nematic flow~\cite{guillamat2016probing}. Other techniques involve the use of more structured environments that can be controlled by external fields and themselves patterned with defects, such as depositing active nematic films onto bulk smectics~\cite{guillamat2017taming,guillamat2016control} (Fig.~\ref{fig:Fig1}h). When in contact with the active nematic layer, focal conic domains in the smectic, for instance, cause the active disclinations to swirl along circular trajectories~\cite{guillamat2017taming}. \ss{In 3D, active drops in a passive nematic fluid can also entrain and activate the dynamics of passive defects, such as ring disclinations in the surrounding nematic~\cite{guillamat2018active,rajabi2021directional}.} This provides a neat way to rapidly reconfigure and channel flows in the system by rectifying defect motion.

Physical confinement has also been studied, particularly in the disc geometry. \ss{Spiral vortices were first stabilized in bacterial suspensions by confinement~\cite{wioland2013confinement,lushi2014fluid}, though their size was limited ($\lesssim 80-100~\mu$m) by the intrinsic instability toward turbulence. High solvent viscoelasticity can ``calm'' active turbulence and allow larger millimeter-scale bacterial vortices with both coherent and globally oscillating flows~\cite{liu2021viscoelastic}. Polar fluids of colloidal rollers, which lack the hydrodynamic interactions that causes bacterial turbulence, self-organize into macroscopically stable vortex pattern when confined  to circular tracks~\cite{bricard2015emergent} and exhibit novel states of pinned vortices in the presence of  quenched disorder in the substrate~\cite{chardac2020meandering} (Fig.~\ref{fig:Fig1}c).} Microtubule-based active nematics also develop steady circulating flow when strongly confined, giving way to a pair of $+1/2$ defects (as topologically required) nucleating at the boundary and orbiting around each other~\cite{norton2018insensitivity,opathalage2019self}. For large disc sizes, more defects nucleate and unbind in the bulk, degenerating into active turbulence (Fig.~\ref{fig:Fig1}a). Laterally confined nematics support shear states into which $+1/2$ defects are continually injected, as they swim around each other in a dancing fashion~\cite{shendruk2017dancing,hardouin2019reconfigurable}. While experiments and continuum simulations agree well in many regards, recent work has noted discrepancies in predictions related to defect nucleation and steady flow states in confined nematics~\cite{opathalage2019self}, suggesting more theoretical work is required.

A third way is to confine active fluids using curved substrates. Curvature frustrates order and often necessitates defects~\cite{nelson2002defects}, providing an exciting avenue to pattern active fluids. Nematics assembled onto spherical vesicles  \mcm{must accommodate a net topological charge $+2$, as required by the Gauss-Bonnet theorem~\cite{carmo1992riemannian}. In equilibrium, for equal bend and splay elastic constants, the configuration lowest in energy corresponds to four $+1/2$ defects located at the corner of a tetrahedron inscribed by the sphere~\cite{nelson2002toward,shin2008topological}. In \emph{active} nematic vesicles the four defects are motile and oscillate coherently between two equivalent tetrahedral configurations, driving spontaneous cell-like shape oscillations  (Fig.~\ref{fig:Fig1}b)~\cite{keber2014topology,zhang2016dynamic,khoromskaia2017vortex}.} On a torus, the spatially varying positive and negative Gaussian curvature causes defects to unbind, attracting disclinations to regions of matching sign curvature~\cite{ellis2018curvature}, thereby filtering them by charge. Polar fluids on a sphere instead form vortices at the poles and a distinctive polar band that concentrates near the equator as a result of active advective fluxes pushing material toward the equator~\cite{sknepnek2015active,shankar2017topological}.

\subsection{Biological relevance of topological defects}
One of the most exciting developments has been the recent characterization of topological defects in living tissues, bacterial colonies and even in multicellular organisms viewed as active materials. Elongated cells can form ordered liquid-crystalline textures interrupted by $\pm 1/2$ disclinations. In confluent epithelial tissues~\cite{saw2017topological} and dense cultures of neural progenitors~\cite{kawaguchi2017topological}, cells were found to preferentially migrate and accumulate at $+1/2$ disclinations and escape from $-1/2$ disclinations (Fig.~\ref{fig:Fig1}e). This behavior originates from the large distortion of order around the defect which generates strong local active stresses (Fig.~\ref{fig:Fig1}f). These large compressive stresses drive cell response, and ultimately lead to  cell extrusion and death in epithelia~\cite{saw2017topological}. The structure and role of these active stresses has been confirmed by direct traction force microscopic measurements and via comparison with simulations of  active nematic hydrodynamics.    Similar phenomena have now been reported in \ss{bacterial} systems as well. In growing biofilms, $-1/2$ defects instead provide sites for mound formation and buckling~\cite{yaman2019emergence}, while geometrically patterned $+1$ asters have been suggested to support verticalization~\cite{bacsaran2020formation}. Motile bacteria also display related behaviour; for instance, starved myxobacteria use $\pm1/2$ defects to seed multilayers and cavities to initiate fruiting body formation~\cite{copenhagen2020topological}, while slow moving \emph{P.~aeruginosa} cells outcompete faster mutants that effectively jam at defects and escape into the third dimension~\cite{meacock2021bacteria}.

The role of defects in organizing tissue morphogenesis is another exciting frontier. In the context of developing Hydra~\cite{maroudas2020topological}, topological defects in aligned supracellular actomyosin have been shown to \mcm{correlate with specific morphogenetic processes} in regenerating tissue. Remarkably, along with motile $+1/2$ disclinations, stable $+1$ defects emerge at locations coinciding with the eventual mouth and foot of the organism, thereby defining the body axis well before mophological features appear. Recent \emph{in vitro} experiments with confined myoblasts show tornado-like mound morphogenesis at patterned $+1$ asters that provide sites for growth and cellular differentiation~\cite{guillamat2020integer,blanch2020quantifying}. Active defects have also been found to control the dynamic morphologies of growing cell layers~\cite{comelles2021epithelial}, 2D bacterial colonies~\cite{doostmohammadi2016defect,dell2018growing}, and shape-shifting active shells~\cite{metselaar2019topology}. \ss{We do not yet have general principles to unify these observations, but these results suggest interesting ways in which active defects can constrain or be harnessed to serve diverse evolutionary, developmental, and survival strategies.}
%In contrast to defects in structural order, a distinct, but relevant class of phenomena is posed by defects in active flow fields, \mcm{A question: are these really quantized topological defects or flow vortices?} including from cilliary patterns~\cite{gilpin2017vortex,loiseau2020active} and morphogenetic flows~\cite{streichan2018global} that we don't discuss further due to space constraints.}

On a more subcellular level, the mitotic spindle provides a classic example of a self-organized aster-like defect maintained in constant flux by actin treadmilling and motor activity~\cite{brugues2014physical}. Separate from the above, recent \emph{in vivo} experiments on starfish oocytes also demonstrate excitable biochemical spiral waves and defect chaos in the expression of certain membrane bound signalling proteins~\cite{tan2020topological}. While it remains to be seen what the full implications of topological defects to biology are, it is clear that such an approach is highly productive and at the frontier of active matter research.

\subsection{Experimental advances and defect-based control of active matter}
Beyond elucidating novel phenomena in nonequilibrium physics lies the challenge of harnessing and controlling them. Active matter experiments have made rapid strides in the past decade and are now poised to engineer reconfigurable and programmable materials. A key strategy is to use topological defects as natural motifs to build dynamic structures with organized flow. Biocompatible liquid crystals perfused with swimming bacteria afford simple static control through prepatterned topological defects (Fig.~\ref{fig:Fig1}g) that capture bacteria and direct their collective motion based on topological charge~\cite{peng2016command,genkin2017topological}. A related strategy has also been employed more recently to pattern defects in aligned epithelial monolayers (Fig.~\ref{fig:Fig1}i) grown on structured substrates with strong anchoring~\cite{endresen2019topological,turiv2020topology}.

Modern advances in engineering optical control of biomolecular activity provide a platform to dynamically control and pattern active materials at will~\cite{schuppler2016boundaries,ross2019controlling,zhang2019structuring}. This has been demonstrated in microtubule-kinesin based gels~\cite{ross2019controlling}, where local light activation reversibly self-assembles 3D asters dynamically stabilized by the clustering and polarized motion of motor proteins. These defect structures can be moved and arranged in arbitrary patterns in both space and time (Fig.~\ref{fig:Fig1}j). A similar strategy has also been employed recently in active nematic films to locally pattern active stresses and direct defect motion using spatio-temporal activity gradients~\cite{zhang2019structuring}. Inhomogeneous activity profiles can act as ``electric fields'' and drive the sorting of defects by topological charge~\cite{shankar2019hydrodynamics}, primarily due to the self-propulsion of $+1/2$ disclinations which accumulate in regions of low activity, unlike nonmotile $-1/2$ disclinations.
%(Fig.~\ref{fig:Fig1}j).
This can be exploited to create defect patterns in active fluids and concomitantly design functional materials with targeted transport capabilities. Defect-based control is poised to create innovative active metamaterials in the future~\cite{zhang2019structuring,Norton2020}, possibly facilitated by \ss{data driven techniques}~\cite{li2019data,Colen2020,zhou2021machine}.

%\clearpage
%\newpage

\medskip
\section{Topological band structures in active matter}

\ss{
The topological defects considered so far are robust features of the {\it order parameter} of active media. We now turn to topologically protected waves whose robustness stems from their {\it band structures}. In general, a band structure describes the frequencies 
at which waves (e.g. sound modes) are allowed to propagate as a function of their wavevectors, along with the way the system vibrates at a given frequency.
Frequency ranges in which waves are not allowed to propagate are called band gaps. As a simple example, consider a ring filled with air or any other fluid at rest. The clockwise and counterclockiwse modes corresponding to density waves in the ring are degenerate, resulting in a point where the bands ``touch'' in a lattice of such ring resonators. If a fan that circulates the air in a given direction is added to each ring, the degeneracy is broken (a bit like a spin in a magnetic field). As a result a lattice of such circulators will exhibit a band gap rather than a degeneracy~\cite{khanikaev2015topologically,yang2015topological}.} %\TODO{REFS?}

%By linearizing the models presented in Box 2 about a unperturbed steady-state, we can analyze the fate of small perturbations and determine whether the steady-state is linearly stable or not (see \TODO{REFS} for examples in the context of active matter: Cristina's review/papers, Caussin paper, NR paper, etc.).
%When it is stable, the perturbations are can simply decay, or more interestingly propagate as waves. The band structure dictates the frequencies $\omega$ and the nature of the waves that can propagate at a given wavevector $\bm{q}$ (in other words, the dispersion relations of the waves).
%In general, it also gives the decay/amplification rates of the perturbation ignored in this Box (see Box 4).
%It turns out that these waves can have topological properties: to describe them, we need to introduce a few mathematical objects.

\nocite{kamien2002geometry}
\nocite{Asboth2016}
\begin{figure*}
   \centering
   \includegraphics[width=\textwidth]{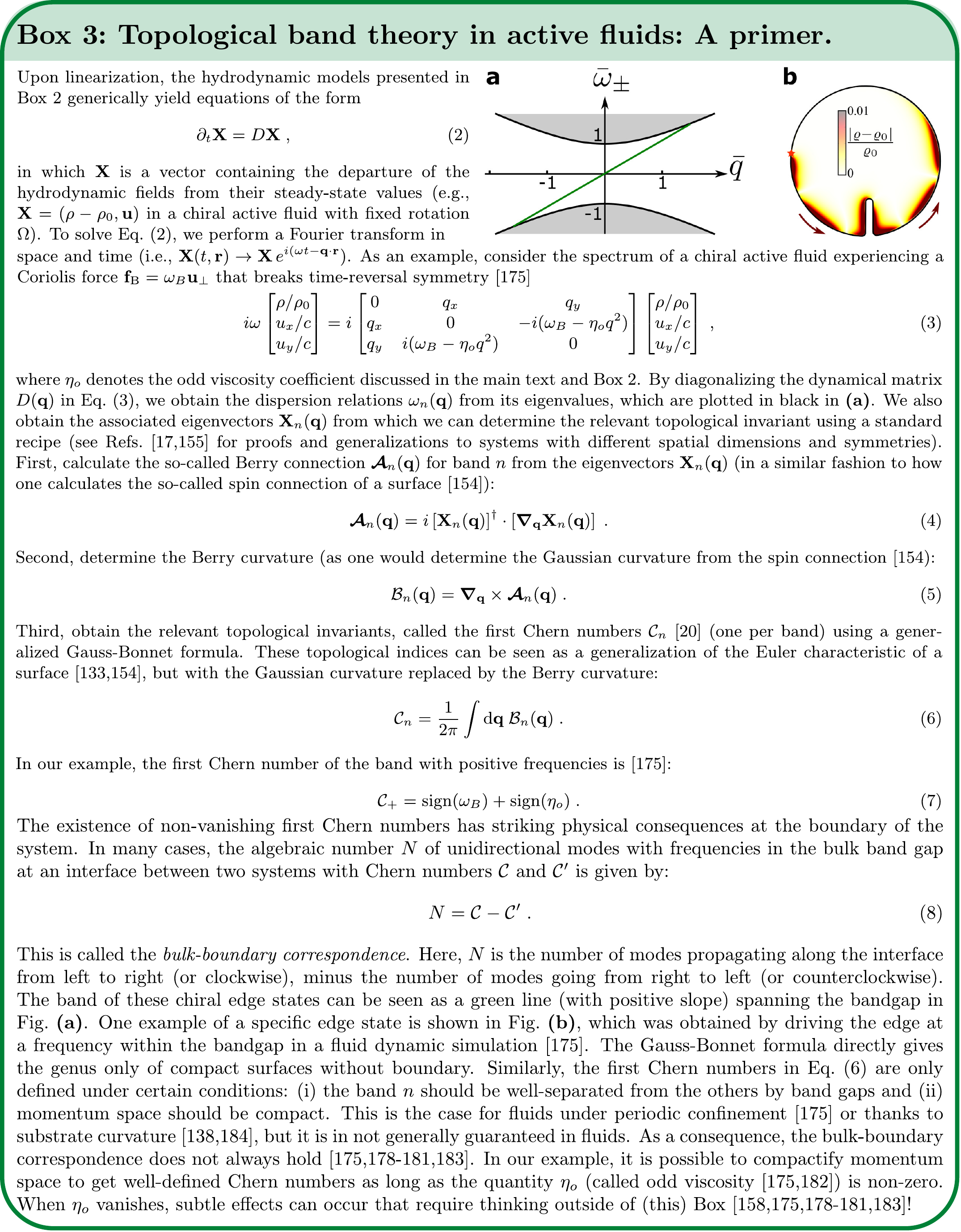}
   \captionsetup{labelformat=empty}
\end{figure*}

\ss{The lattice of circulators exhibits two somewhat surprising properties characteristic of a class of materials called Chern insulators~\cite{hasan2010colloquium,qi2011topological}. First, there are so-called edge states, which have frequencies in the bulk gap---the region where bulk modes do not exist---whose existence is guaranteed by topological invariants in the band structure. Second, these edge states propagate only along the boundaries of the material and only in a single direction (because these states arise from  breaking of time-reversal invariance).}

%The Gauss–Bonnet formula directly gives the genus only of compact surfaces without boundary. Similarly, the first Chern numbers in Eq.~\eqref{chern_number} are only defined under certain conditions: (i) the band $n$ should be well-separated from the others by band gaps and (ii) momentum space should be compact. This is the case for fluids under periodic confinement \TODO{(Souslov circulators)} or thanks to substrate curvature \TODO{(Shankar flocking)}, but it is in general not guaranteed in fluids. As a consequence, the bulk-boundary correspondence doesn't always hold \TODO{(Souslov topological waves, Silveirinha, Tauber, Volovik, Bal)}. In our example, it is possible to compactify momentum space to get well-defined Chern numbers as long as the quantity $\eta_{\text{o}}$ (called odd viscosity) is non-zero. When it vanishes, more subtle things occur that are outside of the scope of this Box.

%\begin{figure*}
 %   \centering
  %  \includegraphics[width=\textwidth]{Box3.pdf}
   % \captionsetup{labelformat=empty}
%\end{figure*}

\begin{figure*}[]
	\includegraphics[angle=0,width=\textwidth]{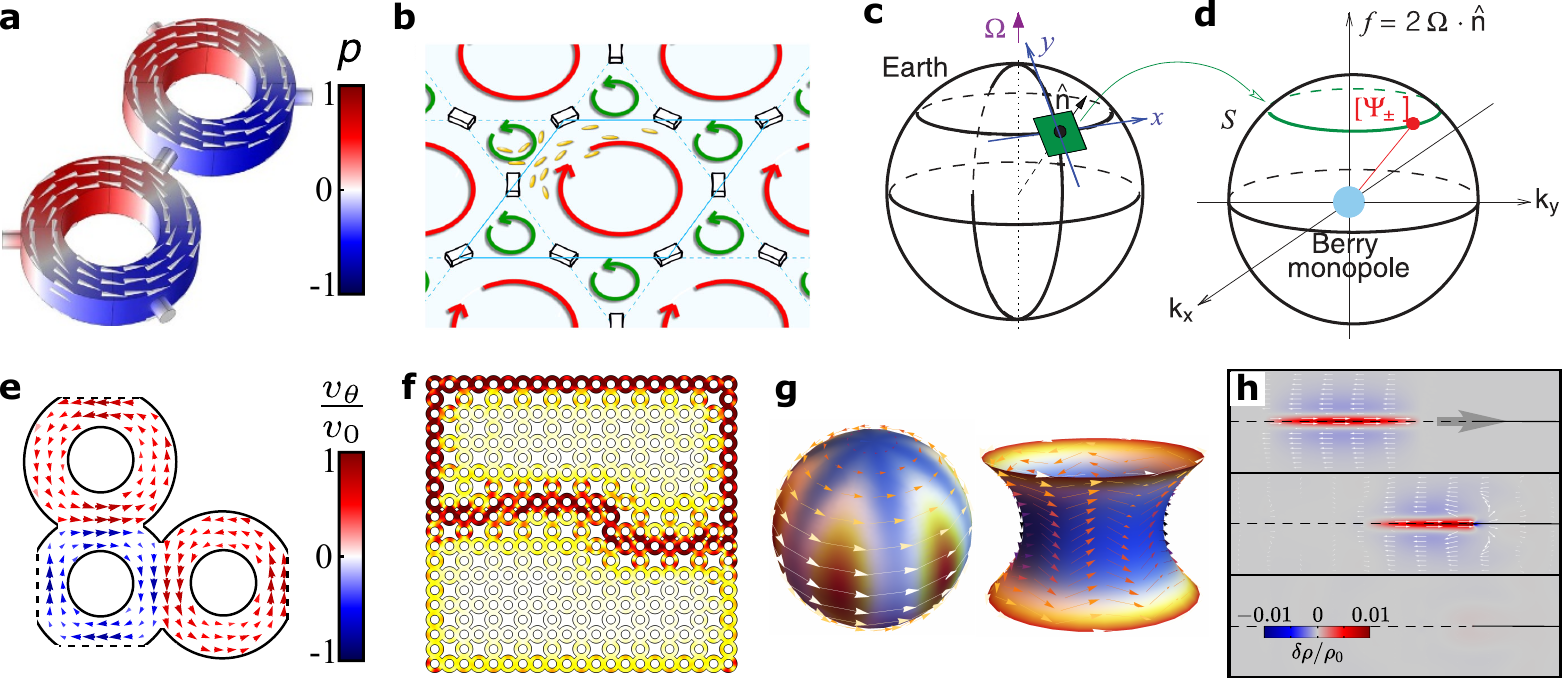}
	\caption{\label{Fig4} %\setstretch{1.75} 
Topological edge states in fluids far from equilibrium. 
(a)	A topological material based on air driven within circulators in which each unit cell has a net vorticity~\cite{khanikaev2015topologically}.
(b)	A polar active fluid with topological edge states but no net vorticity per unit cell~\cite{sone2019anomalous}.
(c—d) Topological waves on the scale of the Earth in which the equator serves as a boundary along which geophysical waves are topologically protected by a Berry monopole due to the Earth’s overall rotation~\cite{delplace2017topological}.
(e—f) Phenomenology of topological edge states in a polar active fluid confined inside a Lieb lattice of annuli~\cite{souslov2017topological}.
Steady state flow~(e) and topological density excitations at edges and interfaces of the lattice~(f) are shown. 
(g) Such topologically protected states arise at the equator in a polar active flock confined to the surface of either a sphere (left) or a catenoid (right)~\cite{shankar2017topological}. \ss{(h) Boundary conditions that do not satisfy bulk-boundary correspondence can be exploited to create perfect absorption: a wavepacket is sent in from the left (top); The wave is absorbed when it encounters a change in boundary conditions (middle), leading to the wavepacket to be entirely absorbed (bottom)~\cite{Baardink2020}}.
}
\end{figure*}

%\subsection{Active media as topological insulators}

\ss{The construction of the topological invariants controlling this robust edge wave propagation can be intuited via a simple mathematical analogy. When the intrinsic (Gaussian) curvature of a toroid is integrated over the entire surface, the result must always equal zero, independent of the precise shape. %This demonstrates that any toroid can be flattened without tearing (i.e., it is topologically equivalent to a square with opposing sides identified).
This is an example of the Gauss-Bonnet theorem~\cite{carmo1992riemannian}, which relates the integrated Gaussian curvature of a closed two-dimensional surface to its genus. This powerful theorem allows us to calculate global topological features of a smooth surface from purely \emph{local} geometric quantities.
Analogously, the topological invariants of band structures are obtained by integrating an abstract curvature describing the geometry of {eigenvalues and eigenvectors} for the linear operator governing wave propagation (Box~3).}

\ss{
Here we review the explanation for how topological band structures result in topologically protected modes at the boundary of a finite sample or at interfaces between systems with topologically distinct band structures.
 At the edge, the Chern number must go from its value in the bulk of the medium to zero (its value outside of the sample). This change cannot happen smoothly because the Chern number is an integer. What happens instead is that the condition for defining a Chern number, namely that the system is gapped, must cease to be valid at the edge, giving rise to confined wave propagation. The edge states resulting from this mechanism exhibit topological protection: they persist even if the properties of the medium are changed, as long as the bulk gap is not closed. As a consequence, the states exist even if defects, obstacles, or sharp features such as corners are present along the boundary. The waves propagate unabated through and around these obstructions without any backscattering: they cannot go back since the edge mode is unidirectional, and cannot penetrate the gapped bulk.} %Nonetheless, they can still be attenuated. 
%so the practicality of this robustness is somewhat dubious.)

\ss{Topological wave propagation is not unique to active media. Besides optical and quantum systems~\cite{hasan2010colloquium,ozawa2019topological}, it can also occur in mechanical systems such as coupled oscillators~\cite{huber2016topological,prodan2009topological,bertoldi2017flexible} as well as simple fluids in circulator arrays~\cite{khanikaev2015topologically,yang2015topological}.
Similar consequences ensue by harnessing active components to induce nontrivial band topology. 
As examples, a topological solid can be realized using ball-and-spring models with active feedback control~\cite{sirota2020nonnewtonian,sirota2021realtime} or by connecting motorized gyroscopes with springs~\cite{Nash2015,Wang2015,Mitchell2018}. 
The combination of the rotation of the gyroscopes and the geometry of the lattice breaks {time-reversal symmetry (TRS)}, and leads to a mechanical Chern insulator with chiral edge states at its boundary. These persist even when some of the gyroscopes are removed or immobilized. In the next subsections, we illustrate the occurrence of topological waves in two classes of active fluids: polar active fluids composed of self-propelled particles and chiral active fluids composed of self-rotating particles.}

\subsection{Topological states in confined polar active fluids}
Polar fluids naturally break time-reversal symmetry through spontaneous flows, which can be directed by geometric confinement to realize emergent chirality and novel topological edge states.
We begin with the example from Ref.~\cite{souslov2017topological} of a polar active fluid confined in periodic microfluidic channels (see Fig.~\ref{Fig4}e--f).
The channel geometry is composed of coupled rings, each reminiscent of an acoustic ring resonator~\cite{fleury2014sound} (Fig.~\ref{Fig4}a).
{Although the fluid itself is polar and achiral, the annular confinement endows the fluid with a spontaneously broken chiral symmetry, which distinguishes between clockwise (CW) or counterclockwise (CCW) spontaneous flow~\cite{wioland2013confinement,bricard2015emergent,stenhammar2016light,thampi2016active}}. 
For a periodic geometry of rings positioned on a square lattice~\cite{wioland2016ferromagnetic}, time-reversal symmetry is restored on average, because there is an equal number of CW and CCW rings. By contrast, removing a single ring from a 2$\times$2 super cell of the square lattice results in a so-called Lieb lattice {(with three rings per square unit cell in an L-shaped pattern, Fig.~\ref{Fig4}e)}. In this case, time-reversal symmetry is broken on the scale of the unit cell. 

This difference between the square and the Lieb lattices has drastic consequences for density waves. Due to the presence of TRS, the square lattice has band crossings at certain symmetric wavevectors. By contrast, TRS is broken in the Lieb lattice and band gaps open between the bands~\cite{souslov2017topological}. Each of these gapped bands can be assigned a Chern number ${\cal C}_n$ (see definition in Box 3) whose value is generically nonzero, and is controlled both by the chirality of the flow and the geometry of the lattice.
Note that in other lattices (such as a honeycomb lattice), it is also possible to obtain topological states without a net unit cell vorticity~\cite{sone2019anomalous}.
In the limit in which the speed of flow $v$ is smaller than the speed of sound $c$, the penetration depth for the localized mode scales as $c a/v$ (where $a$ is the lattice spacing), approaching $a$ if $v \approx c$~\cite{souslov2017topological}. In contrast to driven fluids~\cite{yang2015topological,khanikaev2015topologically}, where achieving this condition requires moderate Mach numbers or resonances, active fluids afford independent control of flow and sound speed, both of which are typically of the same order in experimental realizations~\cite{bricard2013emergence,geyer2018sounds}. {This leads to well-confined edge modes with penetration depth of order lattice spacing.} This feature may be technologically advantageous in the design of miniaturized sonic waveguides~\cite{souslov2017topological, cha2018experimental,cummer2016controlling}. 

Instead of periodic confinement, it is possible to use substrate curvature to produce topological edge states in polar active fluids~\cite{shankar2017topological} (Fig.~\ref{Fig4}g). Gaussian curvature coupled with mean flow breaks Galilean invariance and generically gaps long-wavelength sound modes that acquire a topological character due to the absence of TRS. On the surface of a sphere, a polar active fluid spontaneously circulates around the equator in a chiral fashion~\cite{sknepnek2015active,shankar2017topological}, experiencing an active  analog of the Coriolis force that changes sign across (and vanishes at) the equator. Passive fluids on a rotating sphere, common in geophysical and atmospheric contexts, also exhibit well-known equatorially localized topological sound modes~\cite{delplace2017topological} due to the inertial Coriolis force (Fig.~\ref{Fig4}c-d). In both cases, the equator acts as a gapless interface between two topologically distinct hemispheres. As a result, density waves in a polar active fluid on a spherical surface exhibit unidirectional propagation and topological protection along the equator \ss{in addition to} the polar fluid flows~\cite{shankar2017topological}. This phenomenon is generic to active flow on any surface with nonzero Gaussian curvature and leads to long-wavelength topological sound modes localized along paths that are both geodesics and flow streamlines (Fig.~\ref{Fig4}g).

\begin{figure*}[]
	\includegraphics[angle=0,width=\textwidth]{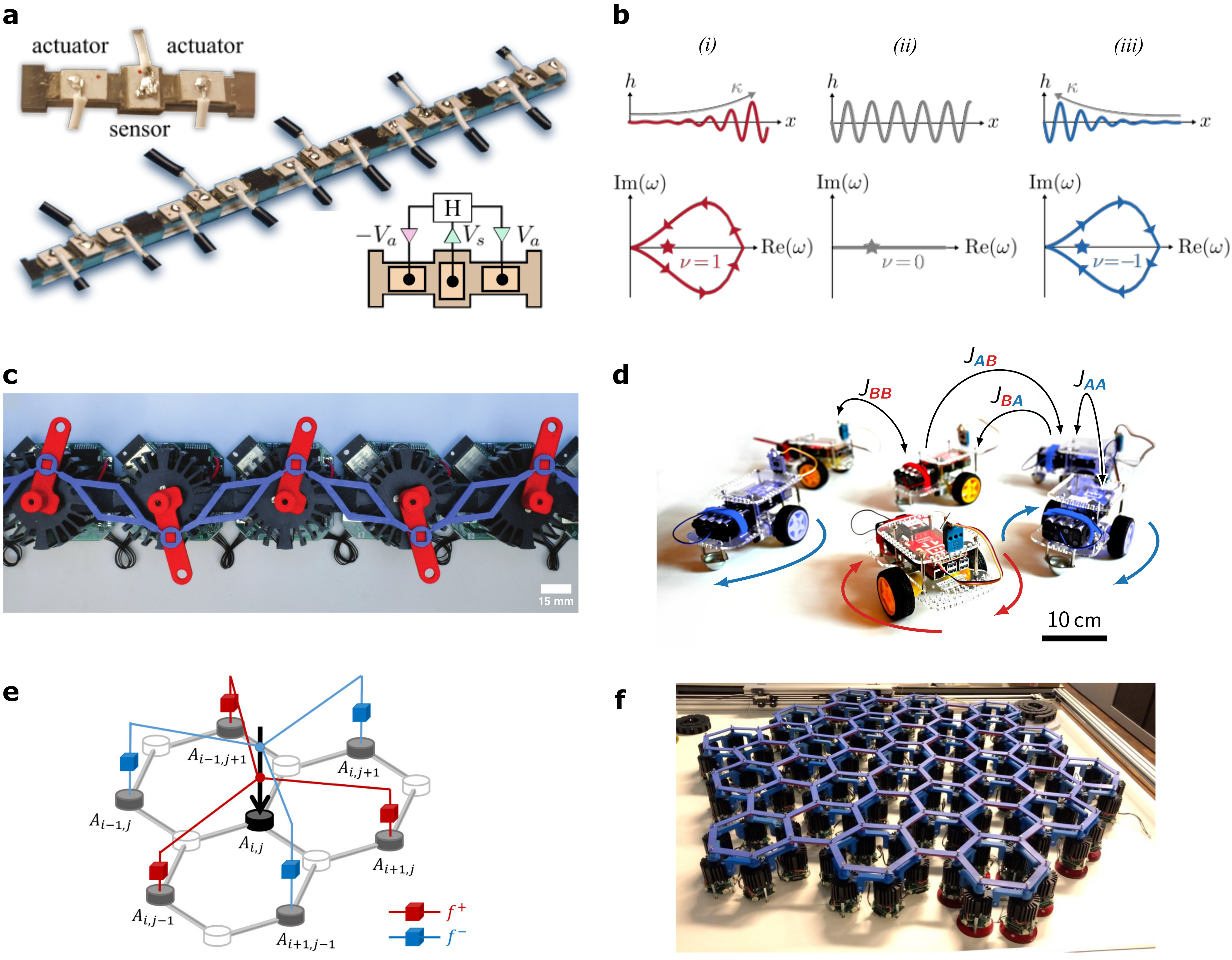}
		\caption{\label{Fig5} %\setstretch{1.75} 
{\it Topology and exceptional points in active and robotic metamaterials}: {\bf (a)} Experimental realization of a self-sensing metabeam with active elasticity. A single unit cell featuring three piezoelectric patches: one that acts as a sensor, and two that act as actuators. Each unit cell has an electronic feedback. {\bf (b)}~Eigenmodes with open periodic boundary conditions with frequency corresponding to the star in the bottom panels. The localization of this eigenmode is determined by the sign of the winding of the energy in the complex plane.  Sketches of the spectrum with periodic boundaries for each of the three cases: \textit{(i)} right, \textit{(ii)} no, and \textit{(iii)} left localization. The arrows indicate directions of increasing wave number $k$. (a--b) are adapted from Ref.~\cite{Chen2020self}. {\bf (c)} A robotic metamaterial composed of an array of sensors and motors (black, with red rods attached) coupled together by soft elastic beams (blue). The motors enforce non-reciprocal interactions in response to forces exerted by neighbors and detected using the sensors. The edge of this one-dimensional metamaterial realizes modes corresponding to a non-Hermitian skin effect. Adapted from Ref.~\cite{brandenbourger2019non}.
{\bf (d)} A swarm of robots programmed to interact as non-reciprocal spins~\cite{fruchart2020phase}. Instead of (anti)aligning like an (anti)ferromagnet, they spontaneously rotate either clockwise or counterclockwise, despite having no average natural frequency. The robotic spins are separated in two populations A (blue) and B (red). The intraspecies exchange interaction $J_{AA}$ and $J_{BB}$ are reciprocal, but the interspecies interactions are not, with $J_{AB} \approx - J_{BA}$.
If self-propulsion is switched on, non-reciprocal flocking models exhibiting exceptional points are necessary to describe the behavior of the robotic swarm. A time-dependent phase with spontaneously broken chiral symmetry emerges in which self-propelled robots run in circles, either clockwise or counterclockwise, despite the absence of any external torque, adapted from Ref.~\cite{fruchart2020phase}. \ss{{\bf (e)} A mechanical lattice whose cells are subjected to active feedback forces processed through autonomous controllers can be programmed to generate desired local response in real time including topological edge states,  reproduced from Ref.~\cite{sirota2020nonnewtonian}. {\bf (f)} A robotic metamaterial is built using a honeycomb lattice with robotic joints whose angular deflections are asymmetrically coupled. It exhibits an odd elastic modulus coupling the two shear modes (while maintaining conservation of both linear and angular momentum) and non-Hermitian skin modes (courtesy C. Coulais).}
}
\end{figure*}

\subsection{Topological waves and odd viscosity in chiral active fluids}
In the above examples of polar active fluids, band-structure topology emerges from the spatial environment that the fluid inhabits. Activity primarily serves to break time-reversal symmetry, which allows for non-zero Chern numbers. On the other hand, chiral active fluids exhibit topological states even in the absence of structured confinement (Ref.~\cite{souslov2019topological} and Box 3). In this case, activity itself endows the fluid with chirality and mesoscopic lengthscales,
leading to {topologically} protected edge states.

For topological states in chiral fluids, activity needs to simultaneously play two distinct roles: (1) breaking time-reversal symmetry and (2) creating a mesoscopic lengthscale in the fluid response. 
Consider a bulk chiral active fluid inside a disk. The fluid will spontaneously rotate due to the balance between local torques arising from self-rotating consituents and dissipation, with rigid-body rotation (having angular velocity $\omega_B$) being the preferred steady state over a broad parameter 
range~\cite{van2016spatiotemporal,soni2019odd,souslov2019topological} (also see Box 1). This rigid-body rotation not only breaks time-reversal symmetry due to flow, but also opens up a band gap around zero frequency in the fluid bulk.
The origin of the band gap is \ss{once again} rooted in the breaking of a basic symmetry of classical hydrodynamics: Galilean invariance (i.e., constant boosts in velocity leave the system unchanged). Rigid-body rotation breaks Galilean invariance by having a fixed rotation axis, and leads to the presence of a band gap. While polar fluids require confinement to generate rotation~\cite{wioland2013confinement,bricard2015emergent,souslov2017topological,shankar2017topological}, chiral active fluids do so intrinsically in the bulk~\cite{van2016spatiotemporal,soni2019odd,souslov2019topological}.

As we explain in Box~3, the Chern number is not always well-defined in fluids~\cite{souslov2019topological,silveirinha2019proof,Volovik1988,tauber2020anomalous,tauber2019bulk}, because the acoustic bands of a fluid are defined on the plane of wavevectors $\b{q}$ which is a non-compact space. This is in contrast with lattice systems, in which the wavevectors are only defined modulo reciprocal lattice vectors (they form a torus called the Brillouin zone, which is compact). \ss{In the example described in Box 3, it is possible to replace the plane by a sphere because of the presence of a dissipationless viscosity called odd viscosity ($\eta_o$, also see Box 2) that can acts as a short distance regularization~\cite{banerjee2017odd,soni2019odd,souslov2019topological,bal2019continuous,tauber2020anomalous}.
A striking feature of chiral active fluids is that the Chern number in Eq.~(7) in Box 3 can change without closing the band gap \footnote{\ss{A change in the Chern number without a band gap closing is also possible in polar active fluids on curved substrates, when the density and orientational sound speeds become equal, causing Galilean invariance to effectively be restored~\cite{shankar2017topological}. However, this mechanism is unrelated to any short scale regularization.}}.
This jump occurs when $\eta_o$ changes sign. The band gap does not close because its width is determined solely by the rotation rate $\omega_B$, but the Chern number in Eq.~(7) still changes. 
In this unusual topological phase transition, the hydrodynamic theory breaks down at short scales as the penetration depth of one of the edge modes goes to zero, and the transition can proceed without band inversion or band gap closure~\cite{souslov2019topological}. 
For the same reason, topological continuum theories also allow for violations of the bulk-boundary correspondence~\cite{souslov2019topological,silveirinha2019proof,tauber2019bulk,tauber2020anomalous,Volovik1988,bal2019continuous,Bal2019b}. \ss{This violation can in turn be exploited to construct waveguides that perfectly absorb a mode in the presence of dissipation (see Fig.~\ref{Fig4}h)~\cite{Baardink2020}.}}

\begin{figure*}
\centering
\includegraphics[width=\textwidth]{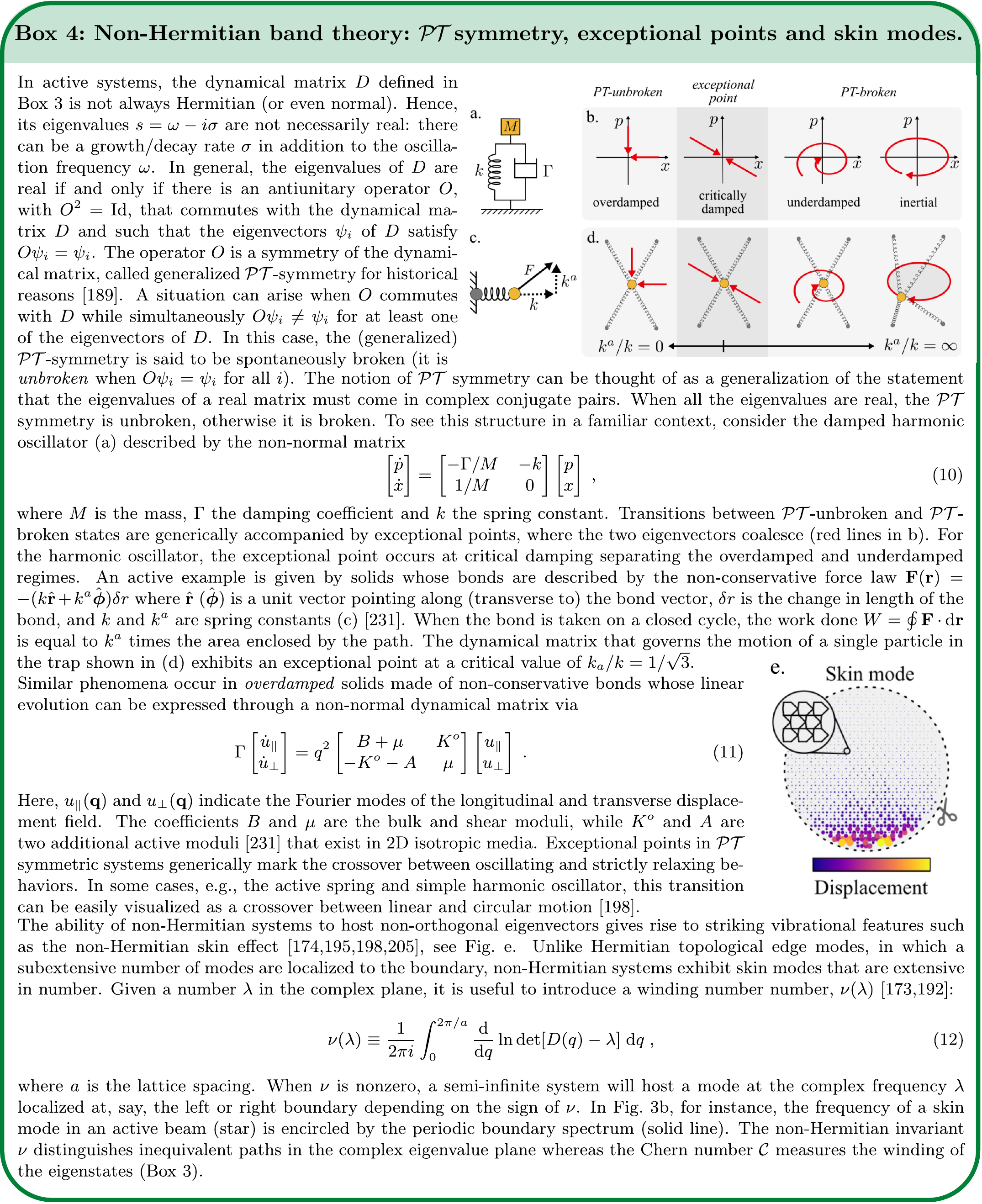}
\captionsetup{labelformat=empty}
\end{figure*}

\ss{\subsection{Non-Hermitian band theory in active media}
In the previous discussion of topological band theory, we have tacitly assumed 
%without any justification 
that the dynamical matrix $D$ %controlling the linearized dynamics 
is Hermitian, i.e., $D = D^\dagger$ (or equivalently, perfectly anti-Hermitian, i.e., $D = - D^\dagger$).
When this assumption does not hold, the band structure can be literally more complex: the eigenvalues are neither purely real nor purely imaginary. 
The real and imaginary parts correspond, respectively, to the oscillation frequencies and decay rates of the relevant perturbations (or vice-versa).
In addition, we also have assumed that $D$ is a normal operator, i.e., $[D,D^\dagger] = 0$. This is always true if $D$ is Hermitian. When this assumption does not hold, the eigenvectors of $D$ need not be orthogonal to each other, see Box 4.
These two changes require generalizations of topological band theory~\cite{Budich2019,ashida2020non,Hatano1996,Bender1998,Bergholtz2019}. %that were carried out in the context of open quantum systems.
%Note that violations of the normality condition implies that $D$ is not Hermitian, i.e., $D\ne D^\dagger$ (but not vice-versa). 
In quantum mechanics, it is natural to assume that the Hamiltonian $D$ of a \emph{closed} system is Hermitian.
%In this case, an Hermitian Hamiltonian guarantees that the energy eigenvalues are real and the time evolution is probability-preserving (i.e., unitary). 
This restrictive assumption has to be relaxed when studying open quantum systems.
%and a %non-Hermitian generalization of topological band theory is usually needed ~\cite{Budich2019,ashida2020non,Hatano1996,Bender1998,Bergholtz2019}. 
Two broad themes have been explored in this context: non-Hermitian skin modes and exceptional points. Before providing an intuitive explanation of these concepts, we stress that their occurrence in classical systems is generic: 
%and it can even help demystify them. 
there is no a priori reason why the linearized operator $D$ should be Hermitian (or even normal), see Box 4.}

\ss{Consider, as a simple example, the advection-diffusion equation~\cite{Trefethen2005}: 
\begin{align}
\partial_t \rho = - v_0 \partial_x \rho + \eta \partial_x^2 \rho\;,\tag{9}
\label{advect}
\end{align}
whose Fourier transformed dynamical matrix $D(q_x)$ is just a complex number given by $D(q_x) = i v_0 q_x - \eta q_x^2$ (where $q_x$ is the wavevector). Manifestly, $D(q_x)\ne D^\dagger(q_x)$ since it has both a real and imaginary parts. Here, $D(q_x)$ is normal but not Hermitian. Equation~(\ref{advect}) describes the transport of a dye with density $\rho$ in a fluid moving with constant velocity $v_0$. When a boundary is inserted in the fluid, the dye accumulates at one end because of the advection $v_0$. While hardly a surprising conclusion, this is in fact 
a very simple manifestation of the non-Hermitian skin effect: a general phenomenon in which the eigenmodes of a non-Hermitian operator are almost all localized to the edge of the system. An iconic example of non-Hermitian quantum mechanics is the so-called Hatano-Nelson model~\cite{Hatano1996}, which is essentially a quantum version of Eq.~(\ref{advect}). From a classical perspective, the asymmetric hopping of electrons (leading to skin modes localized at the edge) in the Hatano-Nelson model can be simply thought as a biased random walk: if the electrons are more likely to hop to the right than to the left, they will accumulate at the right edge.
As explained in Box 4, the skin effect is characterized by a non-Hermitian winding number~\cite{gong2018topological,Zhang2020,Okuma2020}, a topological invariant distinct from the Chern number discussed in Box 3. }

\ss{In active media (Fig.~\ref{Fig5}a--c), various strategies have been devised  to generate and control the localization of energy using the non-Hermitian skin effect~\cite{Budich2019,ashida2020non,Hatano1996,Bender1998,Bergholtz2019,yao2018edge,Zhang2020,Okuma2020,weidemann2020topological, Lee2019anatomy,scheibner2020non,Rosa2020,Zhou2020non, Ghatak2019Realization,borgnia2020non,schomerus2020nonreciprocal, Chen2020self,Tlusty2021,Yamauchi2020,Palacios2020}.
These approaches mostly rely on breaking a family of symmetries collectively known as reciprocity~\cite{Das2002,Lahiri1997,Uchida2010,Saha2019,Gupta2020,fruchart2020phase,You19767,saha2020scalar}. For example, Ref.~\cite{Ghatak2019Realization} uses a 1D chain of coupled non-reciprocal robots whose motors effectively violate Newton's third law (see Fig.~\ref{Fig5}c). Similar effects have been observed in 1D microfluidic crystals~\cite{Beatus2006} and many other soft and active matter systems in which non-reciprocal interactions naturally emerge as a result of the particles being immersed in a medium or in contact with a substrate or field that provides linear or angular momentum~\cite{Beatus2006, soni2019odd, marchetti2013hydrodynamics, kumar2014flocking,bricard2013emergence,PhysRevLett.120.058002, van2016spatiotemporal,Rosa2020,Ivlev2015statistical,Lavergne2019group,Saha2019,Uchida2010, palacci2013living, toner1995long,Chajwa2020,Kryuchkov2018Dissipative,Yifat2018Reactive,Peterson2019controlling,Morin2015collective,Dadhichi2020Nonmutual,Barberis2016large,Gupta2020,loos2019nonreciprocal,toner1998flocks,Bain2019Dynamic,geyer2018sounds,Bertin2006Boltzmann,Farrell2012Pattern,Mishra2010Fluctuations}. 
In this case, the presence of non-Hermitian skin modes can be rationalized using a simplified heuristic argument. Since the forces (or torques) at the two ends of a non-reciprocal bond are not equal and opposite, there is a net momentum flux across each bond. As a result, momentum or energy accumulates at one of the two ends of an open chain determined by the direction of the force (or torque) imbalance very much like the advected dye described by Eq.~(\ref{advect}).}

\ss{So far, we have explained how violations of linear and/or angular momentum conservation lead to non-reciprocity. Can one achieve similar effects (and the resulting non-Hermitian mechanical response) in a self-standing system not coupled to a substrate or any other momentum source/sink? The answer is yes, for example, when the system is active. To grasp this point, we need to consider a notion of reciprocity that is  distinct from Newton's third law---called Maxwell-Betti reciprocity---which can intuitively be defined as the symmetry between perturbation and response. Let us consider an example in the context of Cauchy elasticity in which the deformation of a solid is described by a strain tensor $u_{ij}$ and the internal forces by a stress tensor $\sigma_{ij}$. 
When a passive or active mechanical system is deformed, the infinitesimal work done is $\dd W = \sigma_{i j}  \dd u_{i j}$. For small perturbations about an undeformed state, the stresses are related to strain deformations through $\sigma_{i j} = C_{i j k \ell} u_{k \ell}$ in which $C_{i j k \ell}$ is the tensor of elastic moduli.
%system's linear response obeys Maxwell-Betti reciprocity, the 
If the stress-strain relation can be derived from a free energy $V = \frac{1}{2} C_{i j k \ell} u_{i j} u_{k\ell}$,
%  which is a normal form of the coordinates (?)
then the stiffness tensor $C_{i j k \ell}$ is symmetric, i.e., $C_{i j k \ell} = C_{k \ell i j}$. 
This symmetry is called Maxwell-Betti reciprocity~\cite{nassar2020nonreciprocity}. }

\ss{When the microscopic forces are not conservative, Maxwell-Betti reciprocity is in general violated because $C_{i j k \ell} \neq C_{k \ell i j}$, and additional active elastic moduli are present, which we refer to as odd elasticity~\cite{scheibner2020odd}.  
As a consequence, the dynamical matrix $D_{i \ell}(\bm{q}) = C_{i j k \ell} q_j q_k$ describing the propagation of elastic waves (see Box 4) is not Hermitian,  allowing for the appearance of wave propagation in overdamped solids~\cite{scheibner2020odd} as well as the presence of the skin effect~\cite{scheibner2020non}.
Figure~\ref{Fig5} shows two realizations of metamaterials with odd elasticity: an active metabeam with asymmetric coupling between shearing and bending~\cite{Chen2020self} and a honeycomb lattice with robotic joints whose angular deflections are asymmetrically coupled~\cite{Corentin_progress}. In both systems, the skin effect has been observed. Its existence has been also inferred from either lattice models~\cite{Ghatak2019Realization, Rosa2020, Zhou2020non, Gao2020} or from continuum equations based only on symmetries and conservation laws~\cite{scheibner2020non, Chen2020self}. 
Besides synthetic metamaterial or colloidal systems~\cite{Chen2020self,Corentin_progress,Bililign2021}, odd elastic responses have recently been reported in living chiral crystals self-assembled from swimming starfish embryos~\cite{Tan2021}. }

\ss{Up to this point, we have mostly focused on the eigenvalues of non-Hermitian dynamical matrices. Yet, the corresponding eigenvectors play as crucial a role. When the dynamical matrix $D$ is not normal, it is not always enough to know the eigenvalues to assess the linear stability of the system. This is because the eigenvectors can fail to be orthogonal to each other with respect to the physically relevant scalar product, e.g., a quadratic potential energy density.
The extreme limit of this failure occurs when two (or more) eigenvectors become collinear: this is called an exceptional point~\cite{heiss2012physics}.
As a consequence, there can be a transient amplification of perturbations that can drastically affect the stability of the system~\cite{Trefethen2005}.
A simple example of exceptional points occurs in the damped harmonic oscillator: the overdamped and underdamped regimes are separated by an exceptional point at critical damping (see Box~4). The same thing occurs in active metamaterials: odd elastic waves start propagating when the eigenvalues of $D$ go from being real (decay) to occurring in complex conjugate pairs (oscillations). This transition can be viewed as a spontaneous breaking of a symmetry called $\PT$ symmetry, see Box 4.
Tangible physical effects occur near exceptional points as a consequence of the non-normality of $D$. 
For instance, there can be a transient amplification of noise and perturbations that can drastically affect the stability of the system~\cite{Trefethen2005,Trefethen1993,Chajwa2020,fruchart2020phase,Hanai2020,Strack2013}.
When the dynamical matrix $D$ is viewed as a differential operator in real space, its normality also depends on the boundary conditions. The skin effect occurs precisely when the eigenmodes of a system are orthogonal Fourier modes for a system with periodic boundaries, but become non-orthogonal localized modes in a semi-infinite or finite system. 
Exceptional points can also mediate unusual phase transitions in active many-body systems with non-reciprocal interactions~\cite{fruchart2020phase,You19767,saha2020scalar}.}

\section{Outlook}
Notions from topology and geometry  offer a new perspective on active matter and provide unconventional tools for classifying the complex behavior of these far-from-equilibrium systems. Although abstract, topological techniques are becoming increasingly commonplace in the study of active systems. The success of topology, just as in equilibrium materials, is rooted in the identification of robust collective degrees of freedom and excitations, which in active matter often acquire dramatic properties. %\as{This burgeoning and fast-paced field is  still at a nascent stage}, \TODO{AAARGH} and this review only scratches the surface of topological active matter. 

Exploring the relevance of topological properties of active systems to biology is  perhaps one of the most ambitious avenues in the field. While defects in tissues have been noted as mechanically active centers of morphogenesis~\cite{maroudas2020topological,saw2017topological,kawaguchi2017topological,copenhagen2020topological}, our understanding of the interplay between active forcing and tissue response in manipulating cell organization is still limited. Integrating biologically relevant mechanisms such as growth, cell differentiation, and mechanotransduction with the physics of active fluids would be crucial to this end. On a different scale, although biofilament-motor assemblies routinely display defects when reconstituted \emph{in-vitro}~\cite{sanchez2012spontaneous,kumar2018tunable,ndlec1997self,edozie2019self,weirich2019self}, the relation of these phenomena to \emph{in-vivo} cortical organization~\cite{brugues2014physical,tan2020topological,gowrishankar2012active} remains mysterious and continues to be an open question. During development at the organ and organismal level, both collective motion and pattern formation are ubiquitous~\cite{lecuit2017morphogenesis,howard2011turing}, which offers an intriguing possibility for the realization of exotic topologically protected states. Recent studies have also suggested a role for topological states in nonequilibrium stochastic~\cite{murugan2017topologically}, excitable~\cite{kotwal2019active,hofmann2019chiral,helbig2020generalized,Ronellenfitsch2020}, and evolutionary dynamics~\cite{Knebel2020} networks. 
One challenge faced by future theoretical work is the robustness of topological states to biologically relevant perturbations. In this regard, more systematic experimental investigations are needed.

Active metamaterials and synthetic active matter offer a broad platform to engineer and apply novel topological states~\cite{needleman2017active,peng2016command}. Controlling and patterning defects using activity gradients has already emerged as an exciting direction of research~\cite{shankar2019hydrodynamics,zhang2019structuring,ross2019controlling}. While the main focus so far has been on two-dimensional systems, topological states in three dimensions offer new possibilities, including more complex defect textures~\cite{duclos2020topological} and topological degeneracies in band structure such as Weyl nodes~\cite{armitage2018weyl,Fruchart2018}. 
Designing synthetic gauge fields~\cite{souslov2017topological,Abbaszadeh2017,chardac2020meandering} and exceptional points in active fluids and solids~\cite{scheibner2020non,scheibner2020odd,fruchart2020phase} would be a powerful strategy to exploit these states for sensing and transport. 
Active flow is already being exploited, for example, to enhance \emph{in-vitro} fertilization~\cite{denissenko2012human,kantsler2014rheotaxis}, and creating topologically protected transport could be a route toward new technologies based on active matter. 

From a theoretical perspective, important questions remain. 
Beyond topological band theory, the role of nonequilibrium noise and nonlinear interactions within active materials remains largely unexplored~\cite{fruchart2020phase,You19767,saha2020scalar}.
As discussed extensively, topological defects in active systems acquire much of their novelty from their dynamics~\cite{doostmohammadi2018active}. Although much work has been done, a detailed understanding of defect-driven phase transitions and exotic defect-ordered states remain elusive. Toward this end, comparing and contrasting active defects with the novel collective dynamics of vortices in open quantum systems and driven dissipative condensates~\cite{altman2015two,alicea2005transition,wachtel2016electrodynamic} promises to be a fruitful endeavor.

\acknowledgments
The work of MJB was supported in part by the National Science Foundation under Grant No.~NSF PHY-1748958 and NSF DMREF program, via Grant No.~DMREF-1435794. MCM was primarily supported by the National Science Foundation under Grant No.~DMR-2041459, with additional support from DMR-1720256 (iSuperSeed). SS is supported by the Harvard Society of Fellows.  VV~was supported by the Complex Dynamics and Systems Program of the Army Research Office under grant W911NF-19-1-0268, by the Simons Foundation and by the University of Chicago Materials Research Science and Engineering Center, which is funded by the National Science Foundation under award number DMR-2011854. SS and AS gratefully acknowledge discussions during the 2019 summer workshop on ``Active and Driven Matter: Connecting Quantum and Classical Systems'' at the Aspen Center for Physics, which is supported by National Science Foundation grant PHY-1607611. The participation of AS at the Aspen Center for Physics was supported by the Simons Foundation. AS~acknowledges the support of the Engineering and Physical Sciences Research Council (EPSRC) through New Investigator Award No.~EP/T000961/1. We also acknowledge illuminating discussions throughout the virtual 2020 KITP program on ``Symmetry, Thermodynamics and Topology in Active Matter'', which was supported in part by the National Science Foundation under Grant No. NSF PHY-1748958. We thank M.~Fruchart, C.~Scheibner, G.~Baardink, and J.~Binysh for inspiring conversations and suggestions.

\end{document}